\documentclass[a4paper,12pt]{spieman}  
\usepackage{amsmath,amsfonts,amssymb}
\usepackage{graphicx}
\usepackage{setspace}
\usepackage{tocloft}
\usepackage{tablefootnote}
\usepackage{multirow}
\usepackage{multicol}
\usepackage{xcolor}
\usepackage{lineno}
\newcommand{\matr}[1]{\mathbf{#1}}

\title{Intrinsic Cross Polarization Ratio maps from all-sky observations with the SKA-Low prototype station Aperture Array Verification System 2}

\author[a,f, *, **]{Giulia Macario}
\author[b]{Giuseppe Pupillo} 
\author[b,c,d]{Gianni Bernardi}
\author[a]{Paola Di Ninni}
\author[a]{Giovanni Comoretto}
\author[a]{Pietro Bolli}
\author[b]{Andrea Mattana}
\author[b]{Jader Monari}
\author[b]{Federico Perini}
\author[b]{Marco Schiaffino}
\author[e]{Marcin Sokolowski}
\author[e, f]{Randall Wayth}
\author[e]{Jishnu N. Thekkeppattu}

\affil[a]{Istituto Nazionale di Astrofisica (INAF), Osservatorio Astrofisico di Arcetri, Largo Enrico Fermi 5,  Firenze, Italy, 50125}
\affil[b]{Istituto Nazionale di Astrofisica (INAF), Istituto di Radioastronomia, Via Piero Gobetti 101, Bologna, Italy, 40129}
\affil[c]{Department of Physics and Electronics, Rhodes University, PO Box 94, Makhanda, 6140, South Africa}
\affil[d]{South African Radio Astronomy Observatory, Black River Park, 2 Fir Street, Observatory, Cape Town, 7925, South Africa}
\affil[e]{International Centre for Radio Astronomy Research (ICRAR), Curtin University, Bentley, WA 6102, Australia}
\affil[f]{SKA Observatory, 26 Dick Perry Avenue, Kensington WA 6151, Australia}

\cftpagenumbersoff{figure}
\cftpagenumbersoff{table} 
\begin{document} 
\maketitle

\begin{abstract} 
The low-frequency radio telescope of the Square Kilometre Array (SKA-Low), currently under construction in the remote Murchison shire in the Western Australia's outback, will observe the sky between 50 MHz and 350 MHz with unprecedented sensitivity and stringent requirements for polarization accuracy.
In this work, we investigate the instrumental polarization purity of a SKA-Low prototype station by means of the Intrinsic Cross-Polarization Ratio (IXR) figure of merit. We derive all-sky experimental IXR maps using data from the Aperture Array Verification System 2 (AAVS2). The results are presented at three frequencies within the SKA-Low bandwidth (110, 160, and 230 MHz) with a quantitative comparison between observed and simulated all-sky IXR maps. Our findings show good agreement in IXR map distributions and promising consistency in their radial profiles, meeting SKA-Low’s IXR specification overall. 
This study offers an empirical approach to verifying SKA-Low’s polarization performance using all-sky observations from individual stations and will potentially support the telescope's early science commissioning phase.
\end{abstract}

\keywords{radio astronomy, Square Kilometre Array, phased-aperture-array–telescopes, instrumental leakage, interferometers, polarimetry}

{\noindent \footnotesize\textbf{*} Corresponding author: Giulia Macario,  \linkable{giulia.macario@skao.int}, \linkable{giulia.macario@inaf.it} }\\

{\noindent \footnotesize\textbf{**} \textit{Accepted for publication in Journal of Astronomical Telescopes, Instruments, and Systems on June 4th, 2025}}

\begin{spacing}{1}   

\section{Introduction}
\noindent 
Precise polarimetric measurements of radio signals from the sky are crucial for achieving many scientific goals of modern and future radio telescopes. A key figure of merit (FoM) used to quantify the polarization purity of a radio polarimeter is the Intrinsic Cross-Polarization Ratio (IXR, introduced by Carozzi, Ref. \citenum{Carozzi2011}). This FoM is independent on the coordinate systems and is commonly used to assess the polarimetric performance of low-frequency phased aperture arrays, such as LOFAR \cite{vanHaarLOFAR2013}, MWA \cite{Tingay2013MWA}, LWA \cite{ellings_LWA2009}, and the upcoming low-frequency component of the Square Kilometre Array (SKA-Low) \cite{labate2022jatis}. 
Polarization calibration of such arrays is challenging due to their wide fields of view and the scarcity of polarized calibrators at these frequencies. In the literature, the various works investigating polarization performance using IXR are mostly relying on electromagnetic antenna simulations (e.g., APERTIF, \cite{wijnholds2012} MWA  \cite{sutinjo2013}  and SKA-Low  \cite{BolliDav2022Jatis}) . More recently,  all-sky observations from the LOFAR-LBA Swedish station have been used to estimate the instrumental Stokes parameters across the sky and enable the computation of its IXR\cite{CAROZZI2016}.

SKA-Low is being built at the the Inyarrimanha Ilgari Bundara, the CSIRO Murchison Radio-astronomy Observatory site, in the remote Murchison shire of the Western Australia's outback. It will consist of 512 aperture array stations, each composed of 256 dual-polarized log-periodic antennas \cite{bolli2020}, totaling 131,072 individual receiving elements, providing a collecting area of approximately 0.5 km$^2$\cite{labate2022jatis}. 
Around 50\% of the stations will be located within a circular core of approximately 1 km in diameter, while the remaining stations, grouped in clusters of six, will be positioned along three spiral arms, extending up to a maximum baseline of approximately 74 km \cite{labate2022jatis} within the Inyarrimanha Ilgari Bundara, the CSIRO Murchison Radio-astronomy Observatory site. 
The broadband sensitivity and polarization response of the individual station beams are crucial for assessing the overall performance of the entire SKA-Low telescope, which is key to achieving its scientific objectives. 

AAVS2, the first prototype of a full SKA-Low station with SKALA4.1  antennas, was deployed in 2019 at the Observatory \cite{vanes2020} and has been operational until February 2024 \cite{green2024} . 
It was constructed to verify various SKA-Low station performance metrics and compare them with the system requirements \cite{caiazzo2017} . Since its first light, AAVS2 had been observing the sky and collecting data for different system verification purposes—ranging from digital technology validation to assessments of system stability, calibratability, beamforming, and imaging capabilities \cite{vanes2020, sokolPasa2021, jishnu2024}  . 
Among these efforts, commissioning observations had been used to validate SKA-Low's sensitivity performance, showing good agreement with predictions from electromagnetic simulations and system requirements \cite{sokol2021, macario2022, sokol2022} . Furthermore,  observational evaluations of SKA-Low's polarization performance have begun more recently and are mainly based on full polarization calibration and imaging tests (e.g. Ref. \citenum{ravi2022}) . 
In this work, we present an alternative observational method to derive IXR measurements from single station interferometric all-sky observations making use of available AAVS2 commissioning data. It enables the reconstruction of IXR all-sky maps and their comparison with those obtained from antenna electromagnetic simulations \cite{BolliDav2022Jatis}, allowing to assess the SKA-Low single station polarization response empirically.

The paper is structured as follows: in Sec. \ref{sec:ixr_formalism}, we review the definitions of IXR, summarize the mathematical formalism used in this work, and describe our method for deriving IXR measurements from AAVS2 observations. In Sec. \ref{sec:observ}, we outline the AAVS2 observations and data processing techniques. Section \ref{sec:ixr_obs_maps} presents IXR maps obtained from electromagnetic simulations, describes the IXR mapping algorithm, and provides a comparison between the observed and simulated  maps.  
Results are summarized and briefly discussed in Sec. \ref{sec:discussion}.

\section{Method}
\label{sec:ixr_formalism}

Given a sky brightness distribution $\matr s$, the signal $\matr s'$ measured by a radio dual-polarized instrument in the $(\theta,\phi)$ direction across the sky is given by: 
\begin{equation}
\begin{split}
  \matr s'(\theta,\phi) = [I'(\theta,\phi), Q'(\theta,\phi), U'(\theta,\phi), V'(\theta,\phi)]^T = \\
  = \matr T \, [\matr J(\theta,\phi) \otimes \matr J^*(\theta,\phi)] \, \matr T^{-1} \matr s(\theta,\phi) = \matr M(\theta,\phi) \matr \, \matr s(\theta,\phi) 
\end{split}
\label{eq:sprime}
\end{equation}

where:
\begin{itemize}
\item $(\theta,\phi)$ are the zenith and azimuthal angles in a reference system referred to the array frame\footnote{We note that we use here the $(\theta, \phi)$ coordinate pair to be consistent with the notation introduced, although the standard imaging convention uses the $(l,m)$ cosine directions on the celestial sphere (see \citenum{thompson_book}, pag. 93). This choice does not affect the derivation presented here.}; 
    \item $\matr s$, as a Stokes vector, is 
\begin{equation}
  \matr s(\theta,\phi)=[I(\theta,\phi), Q(\theta,\phi), U(\theta,\phi), V(\theta,\phi)]^T;
\label{eq:s1}
\end{equation}
\item $\matr J(\theta,\phi)$ is the Jones matrix describing the primary beams of the radio dual-polarized instrument at the direction $(\theta,\phi)$; $\otimes$ is the Kronecker product between the Jones matrix and its transposed conjugate;
\item $\matr T$ is the matrix that converts from the instrument frame to the sky one:
\begin{equation*}
  \matr T \equiv \frac{1}{2}
\begin{pmatrix}
1 & 0 & 0 & 1 \\
1 & 0 & 0 & -1 \\
0 & 1 & 1 & 0 \\
0 & -i & i & 0 
\end{pmatrix}\,;
\label{eq:Tmatrix}
\end{equation*}
\item $\matr M(\theta,\phi)$ is the Muller matrix that mixes intrinsic and observed ($^{\prime}$) Stokes parameters.
\end{itemize}

A radio dual-polarized instrument can be affected by \textit{instrumental polarization} leakage $D$  due to the antenna array response that, based on the above formalism, can be defined for unpolarized sky radiation (i.e. featuring $Q, U, V = 0$ and $I \neq 0$) in terms of the first column elements of the Mueller matrix $\matr M$ as:
\begin{equation}
  D(\theta,\phi) =  \frac{\sqrt{M^2_{Q' \leftarrow I}(\theta,\phi)+M^2_{U' \leftarrow I}(\theta,\phi)+M^2_{V' \leftarrow I}(\theta,\phi)}}{M_{I' \leftarrow I}(\theta,\phi)}\, .
\label{eq:leakage}
\end{equation}

\noindent
where the superscript ($^{\prime})$ stands for observed Stokes parameters, 
(i.e. $\matr s'$ Eq. \ref{eq:sprime}), and the $\matr M(\theta,\phi)$ first column's elements are indicated following the notation used in Ref. \citenum{nunhokee2017} to better represent the way intrinsic Stokes parameter $I$ contributes to the observed ones $I',Q',U',V'$: $M_{I' \leftarrow I}(\theta,\phi) = M_{00}(\theta,\phi)$, $M_{Q' \leftarrow I}(\theta,\phi) = M_{10}(\theta,\phi)$, $M_{U' \leftarrow I}(\theta,\phi) = M_{20}(\theta,\phi)$, $M_{V' \leftarrow I}(\theta,\phi) = M_{30}(\theta,\phi)$.

A figure of merit associated to the leakage is the \textit{Stokes (or Mueller) intrinsic cross-polarization ratio} defined as the reciprocal of $D(\theta,\phi)$. In the present formalism, from (Eq. \ref{eq:leakage}), this is:
\begin{equation}
  IXR_M(\theta,\phi) =  
   \frac{{M_{I' \leftarrow I}(\theta,\phi)}}{\sqrt{
   M^2_{Q' \leftarrow I}(\theta,\phi)+M^2_{U' \leftarrow I}(\theta,\phi)+M^2_{V' \leftarrow I}(\theta,\phi)
   }
   }\,.
\label{eq:ixrm}
\end{equation}
We note that, in the absence of instrumental leakage, $IXR_M \rightarrow \infty$; vice versa, when its level is significant, $IXR_M \rightarrow 0$. We refer the reader to Ref. \citenum{Carozzi2011} and Ref. \citenum{nunhokee2017} for a comprehensive derivation and definition of the IXR. 

In Section 3, we will show the predicted and measured all-sky IXR maps for AAVS2. Indeed, being both SKALA4.1 and AAVS2 dual polarized radio instruments, we can estimate their polarization purity degree by applying Equation \ref{eq:ixrm}. As for the predicted $IXR_M$, we will use electromagnetic simulations sufficient enough to calculate the instrument's Jones matrix; while, as for the measured one, we will obtain $IXR_M$ maps across the sky through interferometric Stokes imaging. In order to measure the instrumental leakage, we define a four-element visibility vector between antenna $i$ and $j$ composing the array as four cross-correlation products:
\begin{equation}
  \matr v_{ij} = [v_{xi}v_{xj}^{*}, v_{xi}v_{yj}^{*}, v_{xi}^{*}v_{yj}, v_{yi}v_{yj}^{*}]^T. 
\label{eq:v1}
\end{equation}

The measured visibility vector is related to the sky brightness distribution and the instrument response by the Van Cittert-Zernike theorem (e.g. Ref. \citenum{thompson_book,SynImRA_book2,smirnov2011}):

\begin{equation}\label{eq:v2}
\begin{array}{c}
\matr v'_{ij} = \displaystyle
\iint \matr T [\matr J_i(\theta,\phi) \otimes \matr J_j^*(\theta,\phi)] \matr T^{-1} \matr s(\theta,\phi)  e^{-2 \pi i \nu \frac{(u_{ij} \sin\theta \cos\phi + v_{ij} \sin\theta \sin\phi)}{c}} sin\theta d\theta  d\phi,
\end{array}
\end{equation}
where the integrals are taken over the entire sky hemisphere above the antenna array (\(\theta \in [0, \pi/2]\) and \(\phi \in [0, 2\pi]\)).
After calculating the four cross-correlation products for each baseline of the antenna array (i.e. $u_{ij}$ and $v_{ij}$ for $i,j$ = 1, ..., $N_{ant}$), a 2D Fourier inversion is applied to retrieve the measured images of the four Stokes parameters $\matr s^{'}$ including the instrumental polarimetric response (or leakage):

\begin{equation}\label{eq:v3}
\begin{array}{c}
\matr s'(\theta,\phi) \approx \sum_{i,j} \matr v'(u_{ij},v_{ij}) e^{-2 \pi i \nu \frac{(u_{ij} \sin\theta \cos\phi + v_{ij} \sin\theta \sin\phi)}{c}} .
\end{array}
\end{equation}
Under the assumption that the sky is unpolarized, the measured $IXR^{\prime}_{M}$ to be compared to the simulated one is derived as:
\begin{equation}
  IXR^{\prime}_{M}(\theta,\phi)= 
  \frac{|I'(\theta,\phi)|}{
  \sqrt{Q^{'2}(\theta,\phi) + U^{'2}(\theta,\phi) + V^{'2}(\theta,\phi)}
  }.
\label{eq:l}
\end{equation}

We reiterate that Eq. (\ref{eq:l}) 
is valid under this assumption. 
The sky emission is indeed largely unpolarized. Point sources exhibit a polarization fraction below 1-2\% (e.g. Ref. \citenum{bernardi2013,vanEck2018,vanEck2019,Lenc2016,osullivan2023}) 
and polarized Galactic diffuse emission shows a striking degree of anti-correlation with total intensity 
(e.g. Ref. \citenum{bernardi2009,bernardi2010,bernardi2013, Lenc2016, vanEck2018,vanEck2019}). 
Therefore, the total signal detected in frequency-averaged interferometric images is expected to have a corresponding polarized contribution due to instrumental leakage. 
In the case of wide-field images, like AAVS2, we expect the instrument leakage to be essentially due to the antenna polarization response.  

In light of the above, it is thus possible to derive  
$IXR^{\prime}_M$ by combining the four Stokes images ($I^{\prime}$, $Q^{\prime}$, $U^{\prime}$, $V^{\prime}$) obtained from AAVS2 all-sky observations at each analyzed frequency and in each direction ($\theta,\phi$) across the sky (as described in  Sec. \ref{sec:observ} and Sec. \ref{sec:ixr_obs_maps}).

\section{Observations and data processing}
\label{sec:observ}

For the analysis presented in this work, we used observations carried out with AAVS2 during initial instrumental verification (see also Ref. \citenum{macario2022} and \citenum{sokol2021}). The data were centered at 110.2, 159.4, and 229.8 MHz (hereafter referred to as 110, 160, and 230 MHz for simplicity; see Tab.~\ref{tab:obs}). 
Each observation consists of a series of 0.28-second snapshots (divided into two 0.14-second integrations) repeated every 5 minutes, using a single coarse channel of approximately 925.926 kHz bandwidth (for more details, see  Ref. \citenum{macario2022}). Complex voltages from individual antennas were correlated offline (see Ref. \citenum{Wayth2009,sokolPasa2021}) to generate visibilities with 32 channels, each approximately 28.935 kHz wide, stored in UVFITS format \cite{greisen2017} . No significant RFI was detected in any of the observations. Six malfunctioning antennas were permanently flagged at each frequency, and the first three and last two edge channels were also flagged, reducing the effective bandwidth to approximately 780 kHz.

\begin{table}[htbp]
\caption{Summary of AAVS2 observations used in this work.} 
\label{tab:obs}
\begin{center}       
\begin{tabular}{|c|c|c|} 
\hline
\rule[-1ex]{0pt}{3.5ex}  $\nu_c$  &  Start Time &  Observing Interval  \\
\rule[-1ex]{0pt}{3.5ex}    MHz &  (AWST) & (hours) \\
\hline
\rule[-1ex]{0pt}{3.5ex}  $110.2$  & $2020/04/21$ $20:04:27$& $\simeq12$   \\
\hline
\rule[-1ex]{0pt}{3.5ex}  $159.4$ & $2020/04/08$  $20:01:16$  & $\simeq12$ \\
\hline
\rule[-1ex]{0pt}{3.5ex}  $229.8$  & $2020/04/19$  $20:01:53$  & $\simeq12$  \\
\hline
\end{tabular}
\end{center}
\end{table}

For each frequency, we selected only the snapshots corresponding to local night-time (between sunset and sunrise); the starting Australian Western Standard Time (AWST) and duration selected (each about 12 hours) are reported in Tab.~\ref{tab:obs}. This selection was made to avoid the impact of daytime temperature variations on the receiving system (see e.g. Ref. \citenum{perinijatis2022JATIS}). Additionally, it is ideal for the purpose of this work, as it captures the transit of the brightest part of the Galactic Plane's radio emission across the sky—used as the tracer for IXR measurements (see Sec. \ref{sec:ixr_formalism}).

As described in Ref. \citenum{macario2022} (Section 3), the static delay corrections and the receiver gain equalization applied prior to correlation enable the generation of good quality all-sky intensity images  without further calibration of the visibilities. Furthermore, the observed $IXR^{\prime}_M(\theta,\phi)$, being a dimensionless quantity (as defined in Eq. \ref{eq:l}), does not require an absolute flux density scale. Therefore, we imaged the uncalibrated visibilities to obtain all four polarization $XX^{'}$, $YY^{'}$, $XY^{'}$,  $YX^{'}$ sky maps.\\ 
Imaging was performed using the Miriad package \cite{sault1995} , following an approach similar to that described in Ref. \citenum{macario2022} (Section 3). Visibilities from each snapshot within the selected UT range were Fourier transformed into zenith phase-centered all-sky images (using task INVERT with natural weighting). Each polarization correlation product was imaged separately.
The deconvolution of dirty images was carried out using the Clark CLEAN algorithm (\cite{clark1980}), with a maximum of 200 iterations, to reduce side-lobes from the brightest sources while preventing over-cleaning. Each image covered the entire visible hemisphere at the time of observation, with nearly circular synthesized beams of approximately 4.0$^{\circ}$, 2.8$^{\circ}$, and 1.9$^{\circ}$ at 110, 160, and 230 MHz, respectively. 
For each snapshot, we derived four (pseudo-)Stokes\footnote{We use the term \textit{(pseudo-)Stokes} to indicate that these were computed in the image plane (see Ref. \citenum{HurleyW2014}).} all-sky images by combining the $XX^{'}$, $YY^{'}$, $XY^{'}$, and $YX^{'}$ images as follows:
\begin{equation}
\begin{aligned} 
    I^{'} = \frac{(XX)^{'}+(YY)^{'}}{2}  \\
    Q^{'} = \frac{(XX)^{'}-(YY)^{'}}{2}\\
    U^{'} = \frac{(XY)^{'}+(YX)^{'}}{2}\\
    V^{'} = \frac{(XY)^{'}-(YX)^{'}}{2}
\label{eq:stokes_maps}
\end{aligned}
\end{equation}
These images were then used to compute $IXR^{\prime}_M$ values through Eq. \ref{eq:l} and generate the observed $IXR^{obs\_map}_M$ maps, as described in Sec. \ref{sec:ixr_obs_maps}. 

\begin{figure*}
    \centering
    \includegraphics[scale=0.35]{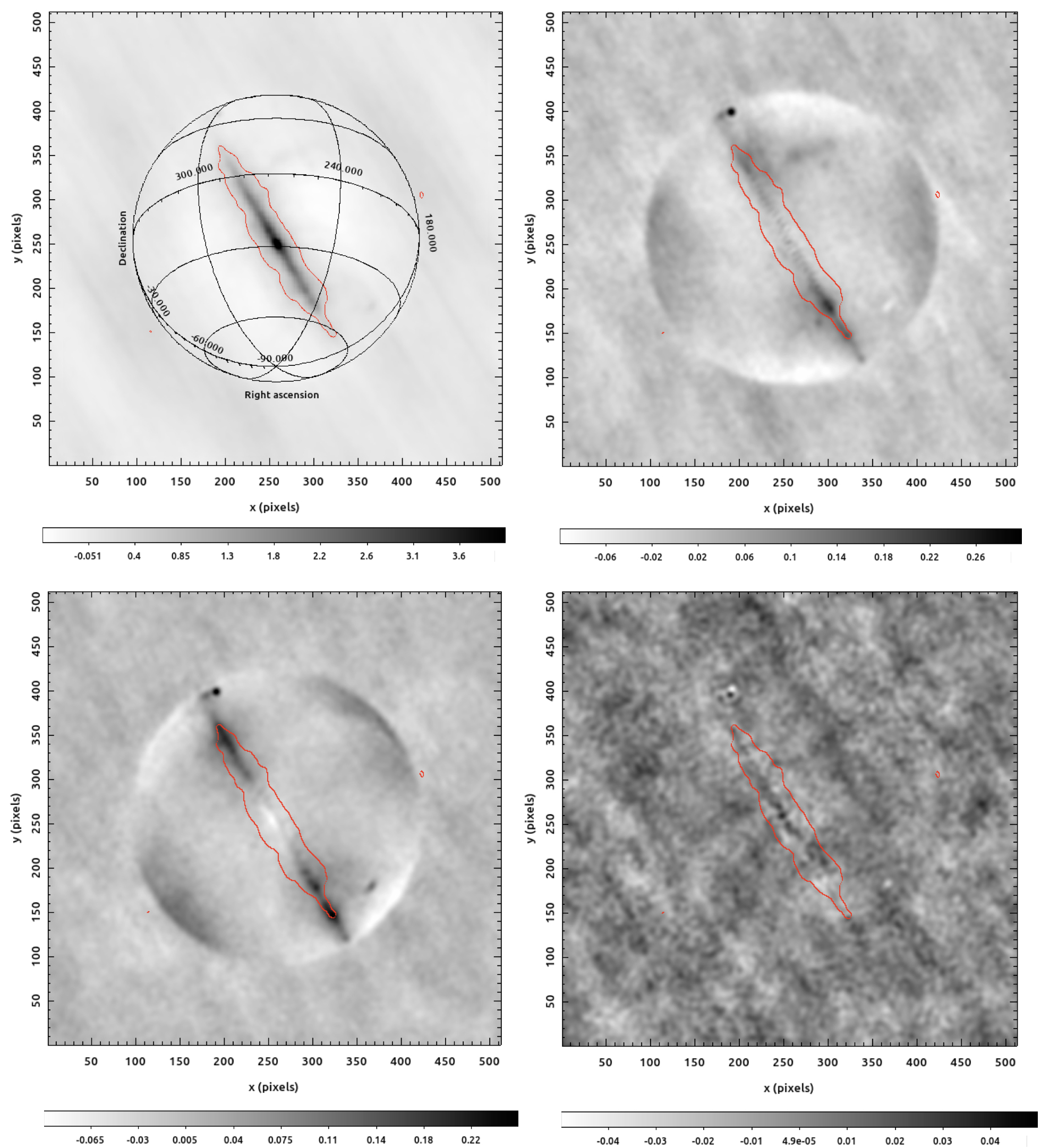}
    \caption{160 MHz (pseudo-)Stokes maps of the snapshot corresponding to the Galactic Plane transit, in image coordinate frame
    : $I^{\prime}$ (top left), $Q^{\prime}$ (top right), $U^{\prime}$ (bottom left), $V^{\prime}$ (bottom right). RA and DEC coordinate grid is over plotted in black to the $I^{\prime}$ image, for sky-coordinates reference. 
    The synthetized beam size is $2.8^{\circ} \times 2.7^{\circ}$, with position angle $-31.1^{\circ}$ (measured from East to North). Grey scale is linear, and colorbar ranges are: $(-0.5, 4)$, $(-0.1, 0.3)$, $(-0.1, 0.25)$, $(-0.05, 0.05)$ counts  beam$^{-1}$ (top left to bottom right, respectively). 
    The red contour overlaid to all images correspond to 5$\sigma_{I^{\prime}}$.}
    \label{fig_stokes}
\end{figure*}

An example of the $I^{\prime}$, $Q^{\prime}$, $U^{\prime}$, and $V^{\prime}$ snapshot images at 160 MHz around the time of the Galactic Plane transit (April 8th, 2020, UT 20:50:57) is shown in Fig. \ref{fig_stokes}. The quality of these images is high, with \textit{Q}, \textit{U}, and \textit{V} signals from the Galactic Plane near zero, as expected (see Sec. \ref{sec:ixr_formalism}). The image noise was estimated by averaging the standard deviation in four square regions (each 100 pixels$^{2}$) located at the image corners: $(\sigma_{I^{\prime}}$, $\sigma_{Q^{\prime}}$, $\sigma_{U^{\prime}}$, $\sigma_{V^{\prime}}) \simeq (0.07, 0.01, 0.01, 0.01)$ counts beam$^{-1}$. The red contour indicates the $5\sigma_{I^{\prime}}$ level, which was also used as signal threshold in the $IXR_M$ mapping algorithm  (see Sec.\ref{sec:ixr_obs_maps}). 

\begin{figure*}[ht!]
   	\includegraphics[width=\textwidth]{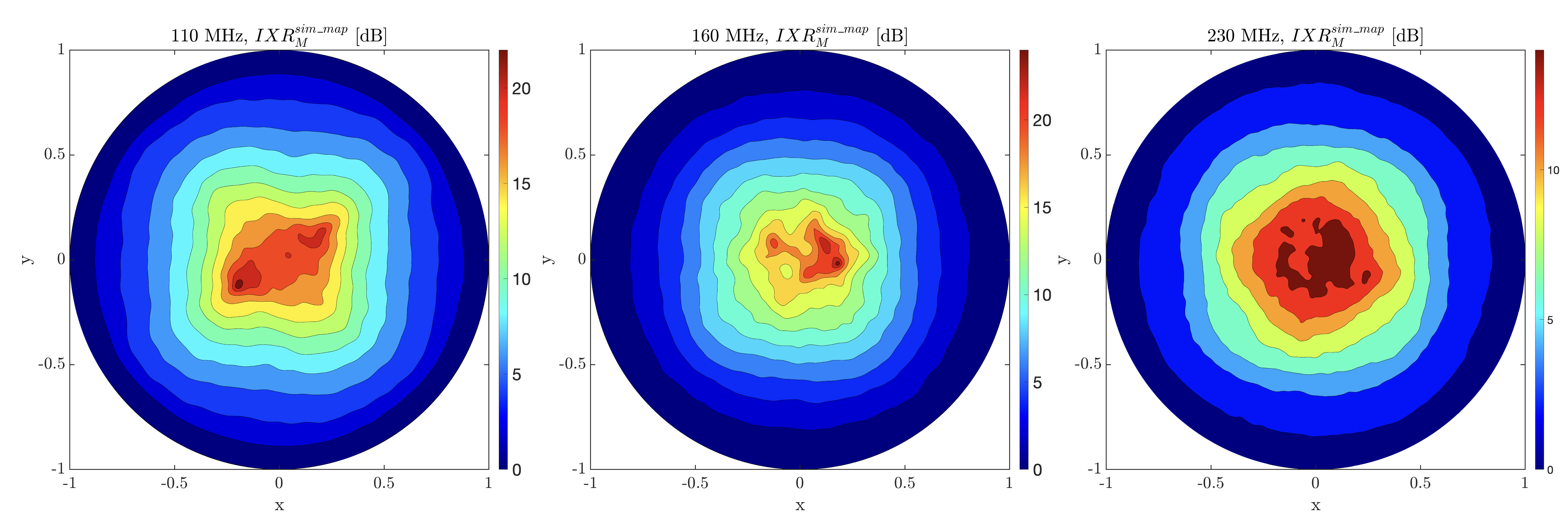}
    \caption{$IXR_{M}^{sim\_map}$ simulated maps in dB, at 110,160 and 230 MHz (from left to right, respectively). The maps are in Cartesian coordinates $\rm (x,y)$: East is towards the right, North is towards the top, and the centre (0,0) corresponds to the projected zenith.}
    \label{fig:fig_ixr_simul}
\end{figure*}

\section{IXR mapping}
\label{sec:ixr_obs_maps}

The simulated $IXR_M$ was calculated using Eq. (\ref{eq:ixrm}), with the Mueller matrix elements obtained from electromagnetic simulations. 
The simulated all-sky maps, hereafter $IXR_{M}^{sim\_map}$, were then generated by phase steering the station beam towards each direction across the sky $(\theta,\phi)$ where the embedded element patterns (EEPs) had been simulated. Note that EEPs were only simulated at discrete pointing directions (0.5 deg resolution in both $\theta$ and $\phi$) and IXR maps were computed only at these points to avoid errors due to interpolation. The $IXR_M$ values were extracted at those specific angles, as detailed in Ref. \citenum{BolliDav2022Jatis}.
Figure \ref{fig:fig_ixr_simul} presents the $IXR_{M}^{sim\_map}$ (in dB) computed for the three frequencies analyzed: 110, 160, and 230 MHz. The maps are displayed in Cartesian coordinates $\rm (x=sin\theta cos\phi,\,y=sin\theta sin\phi)$, with East on the right and North on the top. 
The overall distribution of the maps across the sky shows, at the three frequencies, higher $IXR_M$ values near the zenith, which degrade towards lower elevations.
The IXR values range from 0 to approximately 23.2 dB (110 MHz), 24.9 dB (160 MHz), and 15.9 dB (230 MHz).

Using the method described in Sec. \ref{sec:ixr_formalism}, we developed a procedure that computes the $IXR_M(\theta,\phi)$ (Eq. (\ref{eq:l})) for each direction of the sky by primarily using the Galactic Plane signal detected in the AAVS2 all-sky (pseudo-)Stokes images $I^{\prime}$, $Q^{\prime}$, $U^{\prime}$, and $V^{\prime}$ (see Sec. \ref{sec:observ}, Fig. \ref{fig_stokes}). This produces the observed all-sky map, $IXR_{M}^{obs\_map}$, which is projected onto the same coordinate grid as the simulated $IXR_{M}^{sim\_map}$, allowing for a direct comparison with the maps in Fig.  \ref{fig:fig_ixr_simul}.

In different snapshots, the GP traverses various regions of the field of view, sometimes even partially overlapping. The final observed map is calculated as the weighted average of the IXR maps from individual snapshots, where the weight of each snapshot, $w$, is the inverse of the noise variance in the $I^{\prime}$ map.

The map-making algorithm was implemented in MATLAB \cite{MATLAB}, with a schematic flowchart provided in Appendix \ref{sec:appendix_flow}.
First, the algorithm creates two blank maps: $A$, which accumulates the weighted IXR values, and $W$, which sums the weights. It then iterates over the total number of snapshots ($N$) in the dataset, reading as input the $I_i'$, $Q_i'$, $U_i'$, and $V_i'$ maps (Eq. (\ref{eq:stokes_maps})) 
of the $i$-th snapshot (with $i$ = 1,2,...,$N$). The noise standard deviation (StDev) for each map is calculated by averaging the StDev values estimated in four square boxes (each 100 pixels$^{2}$) at the corners of the image, resulting in: $\sigma_{I_i'}$, $\sigma_{Q_i'}$, $\sigma_{U_i'}$, and $\sigma_{V_i'}$ (see Sec. \ref{sec:observ} and Fig. \ref{fig_stokes}).

For a given pixel in the all-sky image, the $IXR^{\prime}_{M,i}(\theta,\phi)$ (Eq. (\ref{eq:l})) is then computed only if the following conditions on the signal-to-noise ratio are both satisfied: 

\begin{equation}
   I_i'(\theta,\phi)>5\sigma_{I_i'}
\label{eq:cond_I}
\end{equation}

\begin{equation}
  Q_i'(\theta,\phi)>5\sigma_{Q_i'}  \vee 
    U_i'(\theta,\phi)>5\sigma_{U_i'}  \vee 
    V_i'(\theta,\phi)>5\sigma_{V_i'}
\label{eq:cond_QUV}
\end{equation}

The choice of these thresholds is based on the trade-off between achieving a wide coverage in the final IXR map and ensuring the reliability of the signal in the four (pseudo-)Stokes maps (see e.g. Fig. \ref{fig_stokes}), which are both related to the short integration time of the used observations (0.28 s,  Ref. \citenum{macario2022}).
\noindent Along with $IXR^{\prime}_{M,i}(\theta,\phi)$, weights for the maps averaging are computed as $w_i=1/\sigma^2_{I_i'}$.
Next, the weighted quantity $w_i IXR^{\prime}_{M,i}(\theta,\phi)$ is accumulated in the $A$ map, i.e. $A(\theta,\phi) = A(\theta,\phi) + w_i IXR^{\prime}_{M,i}(\theta,\phi)$, 
while the weight $w_i$ is added to $W$ map as $W(\theta,\phi) = W(\theta,\phi) + w_i$.

When the loop over the $N$ snapshots is completed, the full observed map, $IXR_{M}^{obs\_map}$, is computed dividing $A$ by $W$, in the directions where at least one snapshot has met the threshold criteria (Eq. (\ref{eq:cond_I}) and (\ref{eq:cond_QUV})). The remaining parts of the map are left blank. 

This procedure was applied to each AAVS2 dataset (Tab.~\ref{tab:obs}) obtaining the observed IXR maps at the selected frequencies. These maps are presented and described in the next section (Sec. \ref{sec:ixr_sim_obs_req}, Fig. \ref{fig:fig_ixr_comp}, top row). 

\subsection{Comparison analysis}
\label{sec:ixr_sim_obs_req}

\begin{figure*}[ht!]
\includegraphics[width=\textwidth]{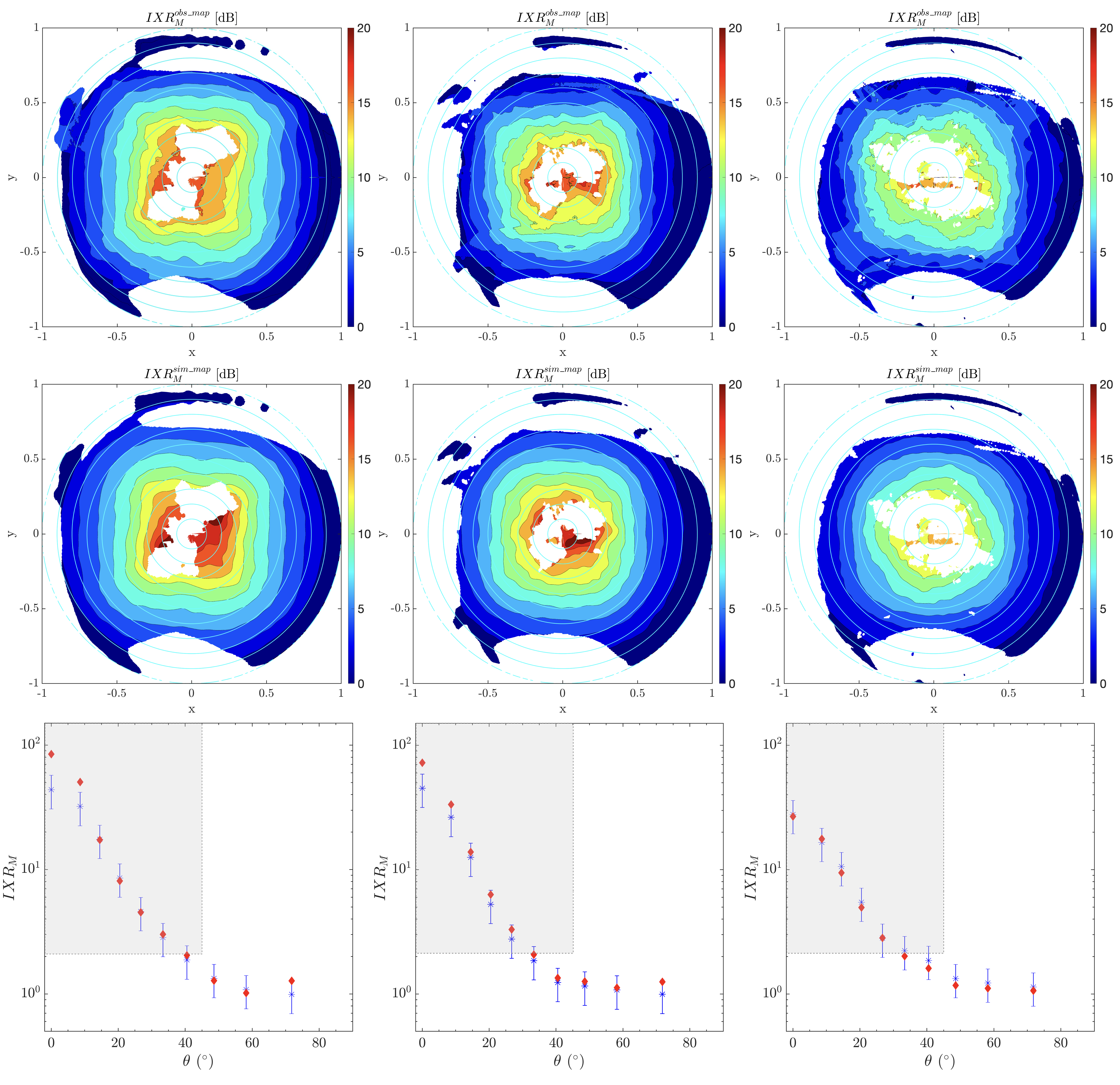}
    \caption{
    Comparison between observed (first row) and simulated (second row) $IXR_M$ maps in dB at 110 MHz (first column), 160 MHz (second column) and 230 MHz (third column). All maps are in Cartesian coordinates $\rm (x,y)$: East is towards the right, North is towards the top, and the centre (0,0) corresponds to the projected zenith. All the color scales range from 0 to 20 dB ($IXR_M$ values from 1 to 100, in linear scale). Simulated maps are the same as in Fig. \ref{fig:fig_ixr_simul}, though masked in the same way as the observed ones for a proper comparison. 
    The bottom diagrams show a comparison between the observed (blue asterisks) and simulated (red diamonds) $IXR_M$ radial profiles, i.e. the average values in concentric annuli (observed and simulated) as a function of the corresponding average zenith angle $\theta$. The uncertainty on the observed $IXR_M$ values is $30 \%$ and is shown by the error bars (see Sec.  \ref{sec:ixr_obs_maps}).
    A logarithmic scale is used to better show the profile.  The shaded gray area represents the SKA-Low requirement, as a reference: $min(IXR_M)$ = 2.12 for $\theta \geq 45^{\circ}$ from zenith ($\theta=0^{\circ}$). See Sec. \ref{sec:ixr_obs_maps}.
    }
    \label{fig:fig_ixr_comp}
\end{figure*}

We present here the comparison between simulated and observed maps. The results are shown in Fig. \ref{fig:fig_ixr_comp}.  
The white regions in $IXR_{M}^{obs\_map}$ maps (Fig. \ref{fig:fig_ixr_comp}, top row) correspond to points where the conditions in Eq. (\ref{eq:cond_I}) and (\ref{eq:cond_QUV}) are not met, and thus $IXR^{\prime}_M(\theta,\phi)$ is not computed (see Sec. \ref{sec:ixr_obs_maps}). For each analyzed frequency, the $IXR_{M}^{sim\_map}$ simulated maps (Fig. \ref{fig:fig_ixr_simul}, central row) were masked in the same way as the $IXR_{M}^{obs\_map}$ observed maps (top row) to allow for a direct comparison. To improve readability, maps are shown in a common range of values, $0-20$ dB. 

Each observed/simulated map pair was divided into ten annular regions, delineated by concentric circles centered at $(x,y) = (0,0)$ with increasing radii, in steps of 0.1 (cyan circles in Fig.  \ref{fig:fig_ixr_comp}). The first region is a circle with a radius of 0.1. The $IXR_M$ values within each region were averaged to produce a 10-point radial profile as a function of the zenith angle, $\theta$. The $IXR_M$ axis is logarithmic to better illustrate the profiles. 
Our $IXR_M$ measurements are affected by various sources of error (e.g., system stability, assumptions about the average EEP during data processing, the accuracy of the gain equalization, and ambient temperature variations) that are difficult to quantify. 
Accurately estimating the uncertainties is complex and beyond the scope of this paper (this may be addressed in future work; see Sec. \ref{sec:discussion}). 
Based on previous studies (Ref. \citenum{sokolPasa2021, macario2022}), we conservatively estimate an error of $\lesssim30\%$ across the entire $IXR_{M}^{obs\_map}$ at all analyzed frequencies (error bars in the radial profiles of Fig. \ref{fig:fig_ixr_comp}).

In particular, the observed and simulated maps show similar distributions across all frequencies analyzed, and there is good general consistency between the observed and simulated $IXR_M$ profiles as a function of the zenith angle (Fig. \ref{fig:fig_ixr_comp}, bottom diagrams). Discrepancies near the zenith ($\theta\lesssim 15^{\circ}$) become more pronounced at lower frequencies, likely due to increased mutual coupling effects between antennas (e.g. Ref. \citenum{BolliDav2022Jatis}), which were not fully taken into account in this analysis (see Sec. \ref{sec:observ}). 
As a reference, the SKA-Low IXR requirement is superimposed on the profiles (gray area in the radial profiles). This requirement is defined in terms of $IXR_J$ in dB and is set at $min(IXR_J) = 12$ dB within the Half Power Beam Width and for observing angles up to $\theta = 45^{\circ}$ off-zenith (Ref. \citenum{caiazzo2017} and subsequent Engineering Change Proposal). In the Mueller formalism  
(Equation 25 of Ref. \citenum{Carozzi2011}), this corresponds to a minimum $IXR_M$ of 3.23 dB (2.12 in linear scale).
Remarkably, this requirement is met down to $\theta \simeq 40^{\circ}$ off-zenith, except at 160 MHz, where it is achieved at even lower elevations, around $\theta \simeq 35^{\circ}$. Note that these interferometric observations are not optimized to validate the SKA-Low IXR requirement, and the requirement itself may still be subject to further review (e.g. Ref. \citenum{dewdney2022}).

\section{Summary and remarks}
\label{sec:discussion}

With the main aim of empirically evaluating SKA-Low single station's polarization response, we have developed and tested a procedure to derive all-sky IXR maps using AAVS2 commissioning observations at 110, 160, and 230 MHz. Under the assumption that the sky emission is largely unpolarized, this method allows us to compute and reconstruct the IXR maps directly from the four interferometric Stokes images of $\sim 12$ hours long snapshot observations. The method is validated by comparing the resulting maps with the simulated ones. Our results demonstrate very good agreement between the observed and simulated maps in terms of distribution across the sky, as well as overall consistency of averaged IXR values in radial profiles. Furthermore, the observed profiles meet the SKA-Low IXR requirement quite well, consistently with previous simulation findings \cite{BolliDav2022Jatis} .

Characterization of SKA-Low's polarization performance is a key aspect of the telescope's test and verification plan and is currently being investigated within the SKA-Low commissioning activities. 
The first four stations, very similar to AAVS2 - with the main difference being the layout\cite{green2024} - , form the \textit{Array Assembly 0.5} (AA0.5), the first operational interferometric array of SKA-Low stations in the SKAO staged delivery plan\footnote{see \href{https://www.skao.int/en/science-users/485/scientific-timeline}{SKAO construction schedule}}, which is currently being commissioned\footnote{see \href{https://skao.canto.global/s/M8159?viewIndex=0}{SKA Phase 1 Construction Proposal} for more details}. 
It might be useful to repeat this analysis during the commissioning phase, using deeper observations from individual SKA-Low stations and covering a broader range of frequencies across the SKA-Low bandwidth. Moreover, since AAVS2 was decommissioned in late 2024, repeating this analysis on SKA-Low single station data could be valuable for improving the method (e.g., placing lower limits on station IXR within masked regions near the zenith, and providing a more detailed assessment of the sources of errors, as discussed in Sect. \ref{sec:ixr_obs_maps}). This is left for the future, subject to the consideration of SKA-Low science commissioning priorities and SKAO data policies.

Finally, we note that beyond the verification purposes for SKA-Low, our proposed procedure is potentially applicable (with some adaptation) to all-sky snapshot interferometric observations from other compact low-frequency aperture arrays, such as e.g. LOFAR2.0\cite{hut2024} or MWA\cite{Tingay2013MWA} stand-alone stations.

\newpage


\appendix    

\section{Appendix}
\label{sec:appendix_flow}

\begin{figure*}[ht!]
   	\includegraphics[width=\textwidth]{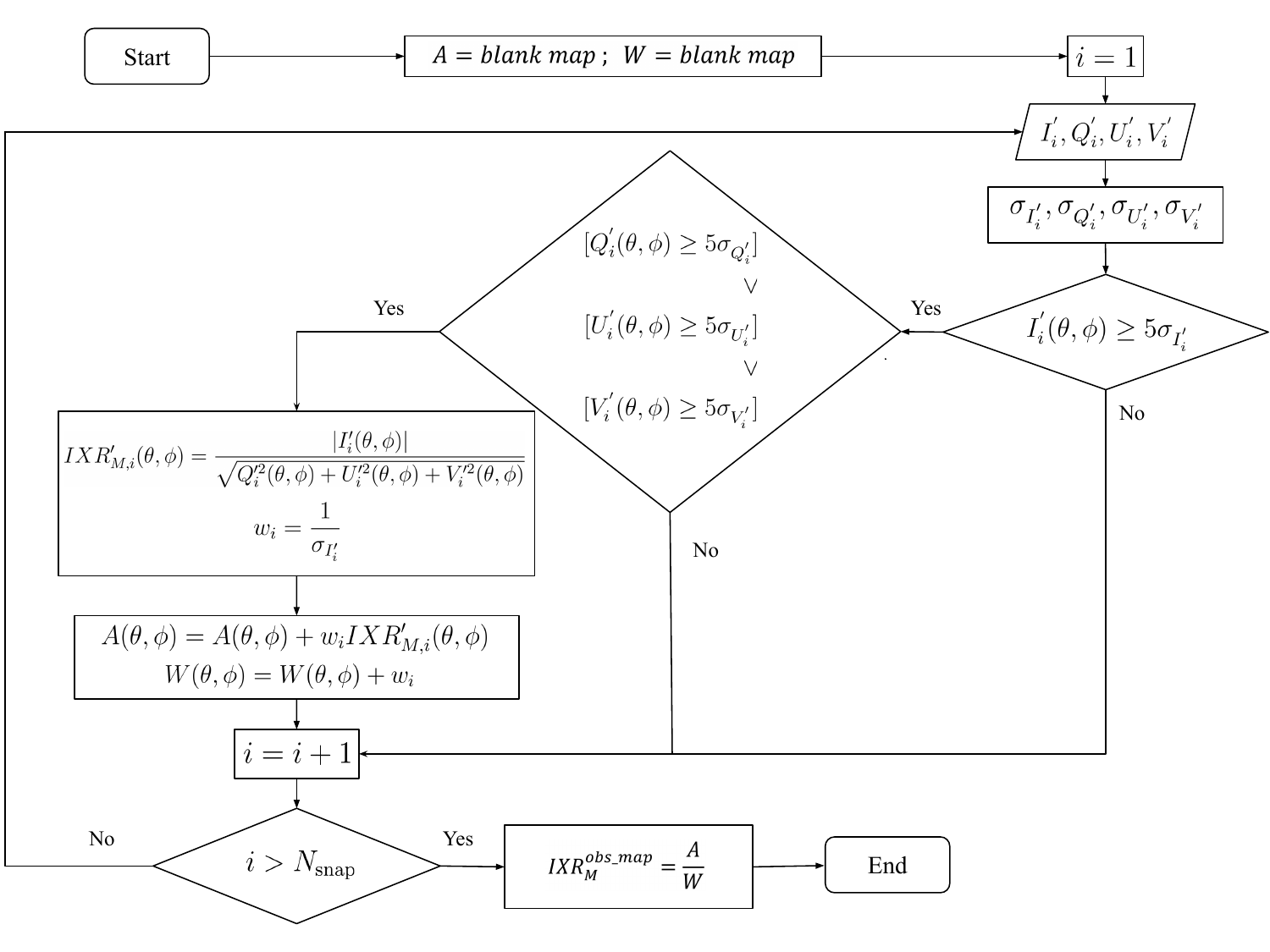}
    \caption{Schematic flow diagram of the algorithm used to obtain the observed maps ($IXR_{M}^{obs\_map}$) analyzed in this work. The description of the algorithm is provided in Sec. \ref{sec:ixr_obs_maps}.} 
    \label{fig:flowchart}
\end{figure*}

\subsection*{Disclosures}
The authors declare that there are no financial interests, commercial affiliations, or other potential conflicts of interest that could have influenced the objectivity of this research or the writing of this paper.

\subsection* {Code and Data Availability} 
The data and codes underlying the findings of this article are currently unavailable to the public, as they are part of the SKA-Low telescope's pre-construction and commissioning phase. However, they can be requested from the corresponding author.


\subsection* {Acknowledgments}

The authors wish to acknowledge the International Centre for Radio Astronomy Research (ICRAR), 
co-founder of the AAVS2 station. AAVS2 is hosted by the MWA under an agreement via the MWA 
External Instruments Policy. We acknowledge the Wajarri Yamaji People as the Traditional 
Owners and native title holders of the Observatory site, paying respects to elders past and present.
We thank the many people, across multiple institutions, who have been involved with the AAVS2 system. 
Particularly, we thank S. Asayama, R. Laing, R. Subrahmanyan for fruitful discussions and 
invaluable suggestions during the development of this work, and G. Kyriakou for having read 
the manuscript and provided comments. This research has used NASA’s Astrophysics Data System. 
The corresponding author acknowledges the use of \href{https://openai.com/chatgpt}{OpenAI’s ChatGPT} Large language model for language proofreading and editing support. The final content has been reviewed and approved by the authors. 



\bibliographystyle{spiejour}   




\listoftables
\listoffigures

\end{spacing}

\end{document}